\documentclass[12pt]{article}%
\usepackage{amssymb}
\usepackage{amsfonts}
\usepackage{amsmath}
\usepackage[nohead]{geometry}
\usepackage[singlespacing]{setspace}
\usepackage[bottom]{footmisc}
\usepackage{indentfirst}
\usepackage{endnotes}
\usepackage{graphicx}%
\usepackage{rotating}
\usepackage{natbib}
\usepackage{footnote}
\usepackage{makecell}
\usepackage{booktabs,threeparttable}
\usepackage{pdflscape}
 \usepackage{multirow}
\usepackage{thmtools}
\usepackage{hyperref}
\usepackage{longtable}
\usepackage{color,soul}

\makeatletter
\def\@biblabel#1{\hspace*{-\labelsep}}
\makeatother
\begin{document}
\bibliographystyle{ei}
\doublespacing

\title{Labor Disputes and Worker Productivity:\\ Evidence from the National Hockey League Lockout\thanks{We would like to thank Peter von Allmen, seminar participants at Union College, and Brian M. Mills for their insightful comments and suggestions. We also thank Dian Jiao and Thobile Nzimande for their excellent research assistance. The remaining errors are ours. \newline \indent \emph{Ge:} Assistant Professor, Department of Economics, Skidmore College, Saratoga Springs, NY. Phone (518) 580-8302, Fax (518) 580-5099, E-mail qge@skidmore.edu. \newline \indent \emph{Lopez:} Assistant Professor, Department of Mathematics and Computer Science, Skidmore College, Saratoga Springs, NY. Phone (518) 580-5297, Fax (518) 580 - 5295, E-mail mlopez1@skidmore.edu.}}
\bigskip
\author{Qi Ge \and Michael Lopez}
\bigskip
\date{June 2015}
\maketitle

\sloppy%
\singlespacing
\textbf{Abstract:}
We implement a propensity score matching technique to present the first evidence on the impact of labor supply decisions during labor disputes on worker productivity in the context of professional sports. In particular, we utilize a unique natural experiment from the 2012-13 National Hockey League (NHL) lockout, during which approximately 200 players decided to play overseas while the rest stayed in North America. We separate the players based on their nationality and investigate the effect of playing abroad on post-lockout player performance. We find limited evidence of enhanced productivity among European players, and no evidence of a benefit or drawback for North American players. The lack of consistent productivity impact is in line with literature in industries with large labor rents, and we propose several additional explanations within the context of professional hockey. Our study contributes to the general understanding of the impact of employer-initiated work stoppage on labor productivity.


\strut


\strut


\pagebreak%
\doublespacing

\section{Introduction}

Workers are highly sensitive to pay adjustments. Unsuccessful bargaining with the employers and subsequent work stoppage can easily demotivate workers and lead to frustration and stress. Unsurprisingly, economists have long been interested in studying the impact of labor disputes on productivity and have explored various forms of labor conflicts and considered different measures of productivity from both public and private sectors. However, while theory predicts that pay reduction or labor disputes can lead to reduced morale which in turn causes shirking and lower product quality \citep{bewley1999}, there is no empirical consensus on the direction of the impact. For instance, \citet{mas2006} finds that lower productivity is associated with disappointing labor arbitration outcomes, \citet{kruger_mas2004} demonstrate a positive relationship between defective products and and previous incidences of strikes, and \citet{chandrasekher2013} shows that working under expired contract as a result of lengthy contract negotiations can lead to implicit productivity decrease in the form of more professional misconduct. On the other hand, \citet{lee_rupp2007} suggest limited support for reduced effort and productivity among commercial airline pilots when facing pay cuts.

Among the industries with labor unrest, professional sports is a highly publicized field that has witnessed over a dozen incidents of lengthy or failed collective bargaining in the past three decades. Almost all of these labor negotiations resulted in player-organized strikes or owner-initiated lockouts that led to the loss of part or the entire professional sports seasons. Since the Major League Baseball (MLB) strike in 1994, lockouts have become the dominant form of work stoppage in professional sports leagues in the event of unsuccessful collective bargaining. Understandably, there is no lack of media coverage on these lockouts\footnote{See \url{http://www.huffingtonpost.com/tag/2012-nhl-lockout/} for a list of news updates on the 2012 NHL lockout}. Meanwhile, professional sports has generated a lot of recent scholarly interest in empirical research because it provides detailed and publicly available information on employee salary, and there are clear measures of labor productivity in the field \citep{Kahn2000}. In terms of professional sports lockouts, economists have studied the impact of lockouts on local economy and employment (e.g. \citealp{coates_humphreys2001}), consumer demand (e.g. \citealp{schmidt_berri2004}), fan substitution (e.g. \citealp{winfree_fort2008}), on-ice infractions (e.g. \citealp{kahane2013}), and competitive balance (e.g. \citealp{zimbalist2000}). Intriguingly, there have not been studies that directly explore the impact of lockouts on professional sports player performance despite the abundant literature on the effect of labor conflicts on worker productivity. On a broader note, despite having grown recently to represent a record percentage of work stoppages in the U.S.\footnote{Greenhouse, S. (January 22, 2012). More Lockouts as Companies Battle Unions. \emph{The New York Times}, A1.}, lockouts received considerably less scholarly attention compared to strikes\footnote{For example, \citet{hall2004} is among the few studies that analyze the effect of a lockout outside professional sports, but it does not evaluate the impact on worker productivity.}. The limited coverage could be attributed to the fact that lockouts were once rare in most industries, and employers have historically been favoring letting employee strike instead of initiating a lockout because of its legal and worker replacement implications\footnote{For example, the employer in the U.S. may only temporarily hire replacements under lockouts and locked-out workers are eligible for benefits. On the other hand, workers on strike may be replaced permanently and may not qualify for benefits.}. With the decline of union representation and increasing empowerment to employers in many industries, it is important to understand the impact of lockouts as an alternative form of work stoppage during labor disputes. Professional sports provides an excellent testing ground for studying characteristics of lockouts due to its numerous recent lockout incidents and readily available sources of productivity data. In addition, professional athletes face the necessity and a variety of options to practice their ``production skills" during the lockout, and this unique feature allows us to explore how player behaviors during the lockout affect post-lockout performance. Our paper thus seeks to not only bridge a gap in the sports economics literature but also provide additional evidence of the impact of lockouts toward the general discussion about labor disputes and worker performance.

In this paper, we study the effect of labor supply decisions under employer-initiated lockouts on worker productivity in the context of the 2012-13 National Hockey League (NHL) lockout. The 2012-13 NHL lockout provides a unique natural experiment to study our topic because during this lockout approximately 200 NHL players chose to play in highly competitive professional hockey leagues in Europe to keep up with their athletic conditioning and receive additional compensation, while others decided to stay in North America and either trained on their own or played in less competitive domestic games with nominal compensation\footnote{Out of the approximately 400 players who stayed in North America during the lockout, 169 of them chose to play in the American Hockey League (AHL). The AHL is considered to be less competitive than the NHL and many of the European leagues because it serves as the primary development league for the NHL and caters to primarily inexperienced young players who are just starting a professional career.}. Because the zero-sum nature of hockey games calls for more effort on and off the ice to stay competitive and European hockey leagues are the most competitive alternative outside the NHL, those players who chose to play in Europe during the lockout were presumably less affected by the stoppage of games and thus more likely to keep their athletic conditioning ready for in-season competition than their peers who chose to stay in North America. By comparing the post-lockout performance between players who played overseas and players who stayed during the lockout, we are therefore able to gauge impact of players' labor supply decisions during the lockout on their productivity. We separate players based on their nationality and consider various productivity measures in professional hockey. We then implement a propensity score matching technique to explore the implication of such natural experiment on player performance in the resumed new season and derive the average treatment effect (ATT) of playing abroad among those who chose to play overseas. We find limited evidence of enhanced productivity among European players who chose to play abroad during the lockout. In particular, we find that for European players, those who chose to play overseas during the lockout had significantly more goals per 60 minutes in the long run (through the duration of the resumed season) than those who did not, although such effect is not apparent in the short run (within a month after the season started). Additionally, there is no evidence of a benefit or drawback for North American players.

Our results suggest that playing overseas during the lockout did not systematically improve post-lockout player performance. Possible explanations for a lack of improved performance from playing abroad include: $(1)$ the lockout may actually motivate \emph{all} players to exert effort to prepare for the new season, regardless of the location and format of preparation; $(2)$ the relatively short duration of the lockout may not allow any significant performance gap to form, especially since timely resolution was already expected based on past lockout experience in the NHL; $(3)$ European hockey leagues are different from the NHL in terms of intensity, rink sizes and rules, which may require significant adjustment from players, especially North American ones; $(4)$ players may not exert full effort while playing in Europe due to weak financial incentives and concerns about potential injuries, and they may also have chosen to play in Europe for non-competitive and non-monetary reasons.

Our paper makes the following three contributions to the literature. First, by identifying the effect of playing abroad on post-lockout productivity, our study is the first paper that provides evidence on the impact of labor conflicts and work stoppage on employee performance in the field of professional sports. Secondly, economists and statisticians have recently started to utilize matching techniques in the field of sports economics (e.g. \citealp{DePaola2011} and \citealp{Krumer2014}), and our paper is among the first to employ propensity score matching to study causal inferences in sports industrial relations. Lastly, this paper contributes to our general understanding of the effect of employer-initiated lockouts, an increasingly popular phenomenon across industries, on labor productivity and provides policy implications.

Our paper is organized as follows. Section 2 provides background information on the 2012-13 NHL Lockout. Section 3 introduces a simple conceptual framework for our study. Section 4 outlines the propensity matching technique employed in the study. Section 5 describes the data used in the study, and present results from the design and analysis phases. We then discuss the empirical results in Section 6 followed by concluding remarks in Section 7.

\section{Background on the 2012-13 NHL Lockout}

Due to disagreement between the owners and the players on revenue sharing schemes and contract structures, the four professional leagues of major sports in the U.S. have witnessed more than a dozen strikes and lockouts in the past four decades\footnote{``Pro Sports Lockouts and Strikes Fast Facts", \emph{The CNN Library}, July 28, 2014.}. A strike happens in a labor dispute when players collectively refuse to play for a number of games while a lockout is the cancelation of part or all of the professional season initiated by team owners. Since the 1994-95 MLB strike that resulted in the cancelation of the remainder of the season, work stoppage in professional sports has since been predominantly in the form of lockouts because lockouts offer a clear bargaining advantage for the owners by transferring most of the financial burden to the players\footnote{A lockout is announced prior to the start of the season, which means players are not being paid for each day they are on the labor dispute with the management and thus shifts much of the economic burden of work stoppage to players.} \citep{lockout2012}. Among the the recent lockouts in professional sports leagues, the NHL became the first professional league in North American to cancel the entire season as a result of the 2004-05 lockout.

Following the season-ending lockout in 2005, players and the NHL reached a 7-year agreement, due to expire by the beginning of 2012-13 season. Despite a strong growth of over 50\% in league's revenue and additional TV broadcast deals during the seven seasons following the lockout, the revenue was generated and kept by a few large city teams and almost half of the teams claimed to have incurred a financial loss in the 2011-12 season\footnote{``NHL Team Values: the Business of Hockey", \emph{Forbes.com}, November 28, 2012.}. On the other hand, players were not as affected by their teams' worsening financial conditions, thanks to the 57-to-43 split on the hockey-related revenue between players and owners. Such disparity in economic well-being between the players and the owners naturally led to strife when the renewal time came for a new collective bargaining agreement. Specifically, owners intended to drastically reduce players' share of the hockey-related revenue while players were reluctant to budge and instead demanded profit sharing among teams\footnote{Owners hoped to reverse the split ratio to 57-43 favoring owners while players were willing to lower their share by only a few percents if there was a new revenue sharing among teams.}. Owners also hoped to increase the free-agency threshold to age 30 as well as ten full years of NHL experience\footnote{Under the 2005 agreement, the age and tenure threshold for unrestricted free agent is age 27 and 7 years, respectively.}. The negotiation started near the end of the 2011-12 season, but no agreement was made before the old collective bargaining agreement expired on September 15, 2012. The NHL called a lockout for all pre-season games on September 19 and progressively canceled all regular-season games up to January 14, 2013 as the two parties continued to disagree on the make-whole issue (i.e. whether to honor contracts from previous agreement) and the length of contracts. Just as the hope for a new season started to dim, the two parties resumed the negotiations with mediation from the Federal Mediation and Conciliation Service and miraculously narrowed their disagreement within days. On January 6, the two parties reached a tentative new collective bargaining agreement to end the lockout, and a 48-game new season started on January 19, 2013 following a rushed week of training camp.

Since the 2012-13 NHL lockout canceled only part of the season, it provides a unique natural experiment to study the impact of labor strife and work stoppage on player productivity \footnote{If the entire season were canceled, the new season would resume following the off-season of the canceled season. The impact of the lockout would be blurred by the fact that players could be more prepared when they had an extra off-season to prepare for the new season.}. While waiting for the settlement to be reached, approximately 200 players began signing contracts with teams in European leagues\footnote{Major European hockey leagues that NHL players played at during the 2012-13 lockout include Kontinental Hockey League in Russia, Hockeyallsvenskan in Sweden, National League A in Switzerland, SM Liiga in Finland and etc.} and a few players also considered signing with the AHL. The 2012-13 NHL lockout thus offers a further identification advantage in studying the impact of the lockout on player productivity because there are players who could not only receive compensation but also stay in competitive conditioning by playing in European leagues while the rest of the players decided to practice on their own or played in less competitive games such as those in the AHL and received nominal or no compensation during the lockout. These features of the NHL lockout allow us to formulate a simple conceptual framework and test whether playing overseas effects future performance. We would not be able to explore such identification strategy for the NBA or NFL lockouts because those lockouts did not cause large outflows of players to temporarily play in other comparable professional leagues.

\section{Conceptual Framework}
We present a simple conceptual framework and provide testable hypotheses in this section. Much of the literature on sports production functions focuses on team performance, and researchers have adopted linear specification \citep{Scully1974}, log-linear functional form \citep{gustfson2000}, CES function \citep{bairam1990}, and stochastic production frontier model \citep{kahane2005}. In our study, we are interested in the impact of lockouts on the performance of individual players rather than teams. And we would argue that the player's morale and effort are likely to be affected during a lockout because of the financial loss as well as the uncertainty about future. Thus, we incorporate effort as a key input in player's individual production function and explore its role in a player's preparedness for the resumed season. Since players were in principle not paid by the NHL or their respective teams during the lockout\footnote{The NHL players' association did issue a lockout pay to all players on a monthly basis during the lockout. The pay was between \$5,000 and \$10,000 per player, which was insignificant compared to players' regular salary.}, players' effort may be driven by additional financial compensation they found elsewhere as well as their expectation about the length of the lockout. Similar to \citet{lee_rupp2007}, we then model individual player's effort following the partial gift model by \citet{Akerlof1982}. We treat effort $E$ as a function of wage offer (during the lockout) $w$ and player's view about the survival probability of the lockout $h$, which further depends on how long the lockout has lasted $\tau$\footnote{Much of the literature on effort focuses on incentivizing effort using specific pay structures. This is not the focus of our study as we intend to use a simple framework to highlight the role of effort and its determinants in players' relative preparedness for the season.}:

\begin{equation} E = E(w, h(\tau)) \end{equation}

\noindent where

\begin{equation*} \frac{\partial E}{\partial w} > 0, \quad \frac{\partial E}{\partial h} < 0, \quad h'(\tau)>0 \end{equation*}

This setup implies that effort is increasing in wage and decreasing in player's perception of the survival probability of the lockout. We thus hypothesize that if the European clubs can offer significant compensation, players who played overseas would exert more effort during the lockout than those who trained domestically\footnote{It is worth noting that, consistent with our framework, the impact of compensation on effort may be small during the recent NHL lockout since it is reported that many NHL players only received nominal compensation (especially compared to their NHL contracts) for their games in Europe.}. As the lockout progressed, our model also suggests that players who were practicing on their own or playing for limited compensation during the lockout would perceive the end of lockout to be less likely. The erosion of trust, uncertainty about professional development and financial stress may depress these players and cause them to exert less effort over time, which also implies the costly nature of effort in our setting.\footnote{There may also exist a hunger effect as the lockout progressed, i.e. as players started to worry about their career or even making a living, they would actually be motivated to put in more effort. This would then make the effect of lockout duration on effort ambiguous.}.

Since the employer monopsony power in sports has exhibited similar effect as found in other occupations such as public school teachers \citep{Kahn2000}, we adopt an individual player production function following the literature on labor strife and the performance of public school teachers (e.g. \citealp{Todd_Wolpin2006} and \citealp{turner2013}):

\begin{equation} Y_{it} = Y(E_{it}, X_{it}, \xi_i) \end{equation}
\noindent where $Y_{it}$ is a performance measure for player $i$ at time $t$, $E_{it}$ is the effort level as defined in Equation (1), $X_{it}$ is a set of individual and team inputs that positively contributes to productivity, including age, number of competitive games played, and etc, and $\xi_i$ is player's endowed ability. Detailed descriptions of $Y_{it}$ and $X_{it}$ in the context of our data are provided in Section 5. Because we posit that effort, morale and experience (number of competitive games played) are positively linked to player performance in our model, we hypothesize a productivity gap between players who play overseas and players who stayed in North America during the lockout due to the financial gains from playing in European leagues as well as additional number of competitive games played\footnote{As mentioned earlier, since Europe affords the most competitive hockey leagues outside the NHL, we assume that games in European leagues are more competitive than AHL games or individual training on one's own.}. Our framework also implies that such productivity gap may widen as lockout progressed due to diverging effort levels.

\section{Matching Technique}

We adopt the notation of the Rubin Causal Model (RCM), initially described by \cite{splawa1990application} for randomized experiments and adopted for observational data by \cite{holland1986statistics}. Let $Y_i$ be a player-specific performance measure for player $i$, $i = 1, ..., n$, where $n$ is the total number of players, and $Y_i$ is measured after the lockout. Let $T_i$ be our treatment variable, $T_i = \left\{0,1\right\}$, indicating whether or not player $i$ played overseas during the latter months of 2012, and let $\boldsymbol{X_i}$ be a set of covariates that are associated with $Y_i$ and $T_i$. The variables in $\boldsymbol{X_i}$ are `pre-treatment' variables; that is, they are measured before a player made the choice to play overseas.

It is a common practice within the RCM to make the stable unit treatment value assumption (SUTVA). SUTVA states that both the treatment assignment of one subject does not impact the set of potential outcomes for other subjects, and also that there are not multiple versions of each treatment. Under SUTVA, the potential outcome for $i$ can be written as $Y_i(T_i=t) = Y_i(t)$, which is the outcome we would have observed had $i$ received $t$.

Our specific estimand of interest is the effect of playing overseas among the players who chose to play overseas, referred to the average treatment effect on the treated, $ATT$, where
\begin{equation} ATT = E[Y(T=1) - Y(T=0)|T=1]. \end{equation}
The $ATT$ is commonly estimated by averaging the difference in potential outcomes among the population that chose to play abroad, $\frac{1}{n_t} \sum\limits_{i=1}^n  \bigg((Y_i(1) - Y_i(0) )*I(T_i=1)\bigg)$, where $I(T_i=1)$ is an indicator variable for whether or not player $i$ played overseas and $n_t$ is the number of players that chose to play overseas. Because only $Y_i(0)$ or $Y_i(1)$ is observed for each player, $ATT$ must estimated from the data by imputing the missing potential outcomes. This issue is referred to as the fundamental problem of causal inference \citep{holland1986statistics}.

To approximate the unobservable potential outcome, the RCM requires the assumption of strong unconfoundedness, which states that (i) $Pr(\left\{Y(0), Y(1) \right\}|T , \boldsymbol{X})$ = $Pr(\left\{Y(0), Y(1) \right\}|\boldsymbol{X})$ and (ii) $0 \textless Pr(T=t|\boldsymbol{X}) \text{ for } t \in \left\{0,1\right\}$ (\cite{rosenbaum1983central}). Under strong unconfoundedness, treatment assignment and the set of potential outcomes are independent given $\boldsymbol{X}$, and as a result, the contrast of outcomes with the same $\boldsymbol{X}$ provides an unbiased estimate of the treatment's causal effect to units with that $\boldsymbol{X}$.

For a multidimensional $\boldsymbol{X}$, individual matching is difficult and often infeasible. Instead, matched methods using propensity scores are popular for estimating the $ATT$ and related treatment effects \citep{rosenbaum1983central}. The propensity score is defined as $r(t,\boldsymbol{X}) = Pr(T=1| \boldsymbol{X})$. Under strong unconfoundedness, if two individuals have the same $r(t,\boldsymbol{X})$ but different treatment assignments, the difference in their outcomes provides an unbiased unit-level estimate of the treatment's causal effect. For a further description of propensity scores with a dichotomous treatment assignment, see \cite{stuart2010matching}.

Before we look at the data, it is important to point out the unclear validity of SUTVA in our example. For North American players, the possibility of language barriers, difficulty of identifying living arrangements, and a less traditional style of play are all factors of playing overseas that, in all likelihood, made playing abroad easier for European players. It is perhaps too strong of an assumption to believe that North American players and European players received the same experience after choosing to play abroad. Moreover, empirical evidence has shown that European players earn relatively favorable wages in the NHL, and that they were more likely to have played abroad during the lockout, implying that there are additional distinctions between the two cohorts \citep{Allmen2015}. In light of the difficulty in making the constant treatment assumption, we make the a priori decision to estimate causal effects separately for each subgroup; one treatment effect for North American players ($ATT_{NA}$), and another for European players ($ATT_E$).

\section{Data and Empirical Results}

To ensure a large enough sample of games in each season on which to compare players, we only considered players that participated in at least half of the season's games in each of the 2011-12 and the shortened 2012-13 seasons. We eliminated the 16 goalies from the NHL that played abroad; goalie performance has been shown to be inconsistent over time \citep{berri2010evaluation}, and so we focus on skaters, for whom there is a larger sample size. This yielded a sample size of 539 NHL forwards and defensemen, 404 of which were from either the United States or Canada (72\%). The remaining group of 125 skaters, for simplicity, will be referred to as the Europeans; these players are primarily from Sweden (36 players, 7\%) and the Czech Republic (31, 6\%), with others also from Austria, Belarus, Finland, France, Germany, Kazakhstan, Latvia, Lithuania, Norway, Russia, Switzerland, Slovakia, Slovenia, and the Ukriane. Player nationalities were extracted from the website \url{<http://www.quanthockey.com>}.

A player's participation in overseas competition was cross-checked using a list published on \url{<www.tsn.ca>}. A much larger percentage of European players (63\%) chose to play overseas, relative to North American ones (17\%), perhaps speaking to the perceived advantages that players from Europe would have upon returning to play near their hometowns.

\subsection{Design phase}

A matched analysis should be separated into two phases; the design phase, which is done without the outcome in sight, and the analysis phase \citep{rubin2001using}. The goal of the design phase is to reduce the bias in the covariates' distributions between those who received the treatment and those who did not. In essence, we attempt to replicate the balance from a completely randomized design, under which those treated and untreated are similar in expectation with respect to observed covariates. While a randomized design also provides the benefit that unobserved covariates are also similar in expectation, like with many other examples, the randomization of players to participate in overseas competition during the lockout would have been infeasible and unethical.

Choice of $\boldsymbol{X}$ should be driven to satisfy the assumption of strong unconfoundedness, and \cite{stuart2010matching} recommends including any pre-treatment variable known to be associated with both treatment assignment and the outcome. Factors that may influence a player's choice to play abroad and his subsequent performance include his overall performance level prior to the lockout, salary, age, and position. Given our desire to estimate treatment effects separately for North American and European players, covariate matrices are distinct for each cohort.

We identified 13 variables for our propensity score model, drawn from familiarity with the sport of hockey and literature written around the time of the lockout\footnote{Including, among others, \url{<http://thehockeywriters.com/nhl-players-go-overseas/>} and \url{<http://www.wsj.com/articles/SB110565436846425695>}.}. These variables are listed in Table \ref{TabX}, along with a brief description. The first ten variables ($Shot \%$, ... , $Time$ $on$ $ice$) represent player statistics from the 2011-12 season, and were each extracted from the website  \url{<www.war-on-ice.com>}.  $Shot \%$ is defined as the percentage of on-ice shot attempts (on goal or missed) taken by a players team when that player was on the ice. Among hockey analytics followers, this variable is known as a players' Corsi percentage. Empirical evidence has found that $Shot \%$ percentage is preferred for predicting a players' future success over statistics like goals, assists, and plus/minus rating, because the latter metrics are more variable \citep{schuckers2011national, macdonald2012adjusted}. Moreover, possession measures such as relative $Shot \%$ are more informative for evaluating the performance of defensive players, whose primary roles on the ice may not come on offense. Remaining variables are drawn from  \url{<www.war-on-ice.com>}, and help account for a players position (defense or forward), his offensive production, and proxies for his aggressiveness, such as hits and penalties.

The eleventh variable in Table \ref{TabX}, $Salary$, represents the player's yearly salary for the 2011-12 season, and was extracted using \url{<www.capgeek.com>}. Finally, we include each player's age as of February 1 of the 2011-12 season, $Age$, taken from \url{<www.hockey-reference.com>}, and we add a squared term for age to account for a possible non-linear trend between $Age$ and the probability of playing overseas.

\begin{table*}[htbp]
\centering
\caption{Covariates}
\begin{tabular}{l l l}
\toprule
\label{TabX}
 Variable & Description \\
\hline
$Shot \%$ &Shot percentage, 2011-2012 season \\
$Defense $ & $\left\{1 \text{ for defensemen, 0 for forwards}\right\} $\\
$Goals$ & Goals per 60 minutes, 2011-12 season  \\
$Assists$ &Assists per 60 minutes, 2011-12 season\\
$Games$ & Games played, 2011-12 season \\
$Hits$ & Hits, 2011-12 season  \\
$Hits$ $Taken$ & Hits taken, 2011-12 season  \\
$Penalties$ & Penalties, 2011-12 season  \\
$Penalties$ $Taken$ & Hits taken, 2011-12 season  \\
$Time$ $on$ $ice$ & Time on ice per game, 2011-12 season \\
$Salary$ &Salary, 2011-12 season  \\
$Age$  & Age, in years \\
$Age$ $Squared$&$Age^2$  \\  \bottomrule
\end{tabular}
\end{table*}

After estimating the propensity score using separate multivariate logistic regression models for each of the the North American and European players, we match subjects based on their estimated propensity score, $\hat{r}(t,\boldsymbol{X})$, by using the Matching package in R statistical software \citep{sekhon2008multivariate}. Given that matching with replacement has been shown to yield sets with lower bias, relative to matching without replacement, we match with replacement using a caliper of 0.10 \citep{abadie2006large}. Further, because our outcome variables include offensive output measured after the lockout, we use exact matching by position, which ensures that defensemen are only matched to defensemen and forwards are only matched to forwards.

Table \ref{TabPScore} shows one example of a matched pair, European forwards Henrik Zetterberg and Daniel Sedin, along with a subset of variables from Table \ref{TabX} and their estimated propensity scores. The two players are similar on many of the variables in $\boldsymbol{X}$, including $Age$ and $Salary$, although Zetterberg posted a higher goal total, lower assist total, and a better $Shot \%$. Additionally, the $\hat{r}(t,\boldsymbol{X})$'s for Zetterberg and Sedin are within 0.01 of one another.

\footnotesize

\begin{table*}[htbp]
\centering \footnotesize
\caption{Example of players matched and selected covariates}
\begin{tabular}{l l l c c c c c c c} \toprule
\label{TabPScore}
Ethnicity & Position & Player & Abroad? & $\hat{r}(t,\boldsymbol{X})$ &$Goals^*$ & $Assists^*$ & $Shot \%^*$ & $Salary^{*}$ &$Age$  \\
\hline
\multirow{ 2}{*}{European} &\multirow{ 2}{*}{Forward} &  H. Zetterberg & Yes & 0.704& 30&37&66.9& 6.10 & 31 	\\
 & 										 &  D. Sedin          & No & 0.695& 22&47&60.0& 6.08 & 31 \\ \hline
\multicolumn{10}{l}{$^*$For the 2011-12 season. Salary taken in millions of dollars} \\
\bottomrule
\end{tabular}
\end{table*}

\normalsize

\begin{figure}[htbp]
\begin{center}
 \includegraphics[width=4.61in,height=3.31in,keepaspectratio]{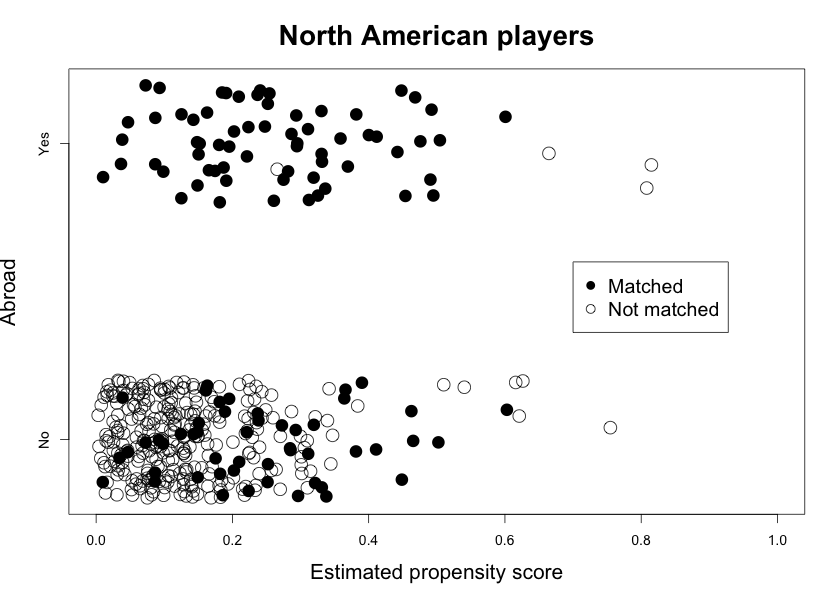}
\caption{Distribution of estimated propensity scores, North American players}
\label{F1}
\end{center}
\end{figure}

\begin{figure}[htbp]
\begin{center}
 \includegraphics[width=4.61in,height=3.31in,keepaspectratio]{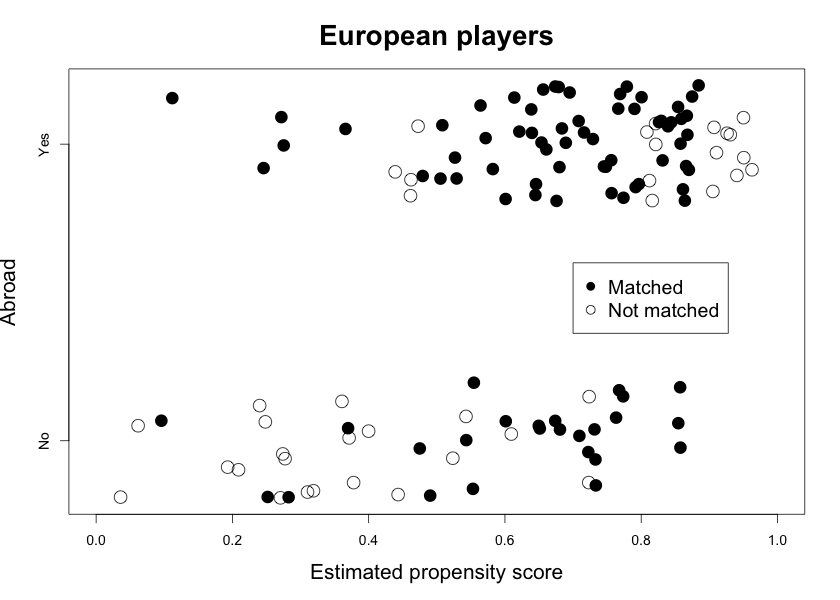}
\caption{Distribution of estimated propensity scores, European players}
\label{F2}
\end{center}
\end{figure}

Figures \ref{F1} and \ref{F2} depict the distributions of $\hat{r}(t,\boldsymbol{X})$ for the North American and European players, respectively. Each of the darkened circles in Figures \ref{F1} and \ref{F2} refer to a player that was matched to someone with the other treatment, and represent the cohort of players that we can generalize to, while subjects not matched are depicted by open circles.

In general, the distributions of propensity scores among those matched appears roughly similar within both the North American and European player cohorts. The $\hat{r}(t,\boldsymbol{X})$'s for North American players, on average, are much lower than European players, and there is a large group of North American players with a very low estimated probability ($\textless 15\%$) of playing overseas. All but four players in the North American cohort that went overseas were matched to a player who did not play overseas; the four subjects not matched did not have a corresponding player at the same position with a similar $\hat{r}(t,\boldsymbol{X})$.

For European players, most $\hat{r}(t,\boldsymbol{X})$'s were between 0.40 and 0.95. There were a handful of players that played overseas with too large of an estimated probability of playing abroad, such that there was no corresponding player to be matched to among the Europeans who chose not play overseas. Including subjects with extreme propensity scores requires extrapolation, and it is often recommended that subjects in this group be discarded \citep{dehejia1998causal}. As a result, these players were not included in our analysis phase.

Differences in the covariates' distributions between treatment groups is commonly assessed using standardized bias \citep{austin2009balance}. Let $\bar{x}_{p1}$ and $\bar{x}_{p0}$ be the sample mean of covariate $p$ among subjects with receiving the treatment and control, respectively, and let $\sigma_{p1}$ be the standard deviation of covariate $p$ in the treatment group. The standardized bias of covariate $p$, $SB_{p}$, is calculated as
\begin{eqnarray} SB_{p}  =  \frac{\bar{x}_{p1}- \bar{x}_{p2}}{\sigma_{p1}} . \end{eqnarray}

\noindent \cite{rubin2001using} recommends that defensible causal statements can be made only if the standardized bias between the treatment and control groups are less than 0.25 for each covariate.

\begin{figure}[htbp]
\begin{center}
 \includegraphics[width=5.91in,height=3.51in,keepaspectratio]{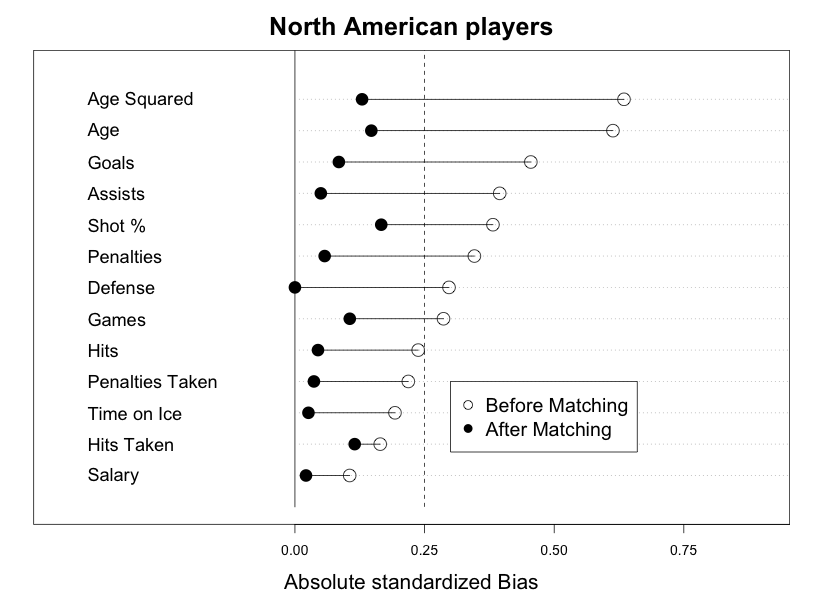}
\caption{Absolute standardized biases, North American players}
\label{F3}
\end{center}
\end{figure}

\begin{figure}[htbp]
\begin{center}
 \includegraphics[width=5.91in,height=3.51in,keepaspectratio]{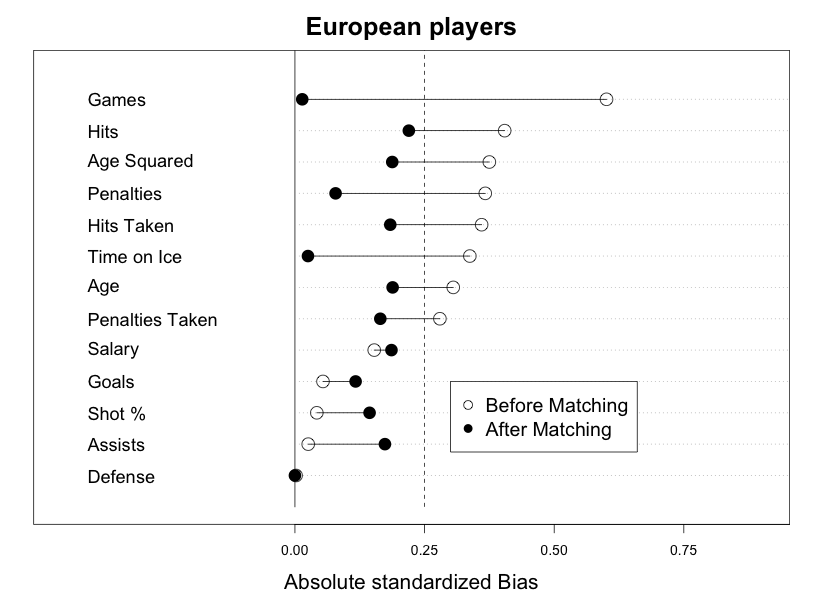}
\caption{Absolute standardized biases, European players}
\label{F4}
\end{center}
\end{figure}

We calculated the standardized bias between players who played overseas and those that did not, both before and after the matching. Figures \ref{F3} and \ref{F4} show `Love' plots of the pre and post-matched absolute standardized biases for both the North American and European cohorts, respectively \citep{ahmed2006heart}.

Of the thirteen standardized biases in each of the unmatched groups of players, eight are greater than 0.25 for each of the North American (in descending order, $Age^2$, $Age$, $Goals$, $Assists$, $Shot \%$, $Penalties$, $Defense$, and $Games$) and European ($Games$, $Hits$, $Age^2$, $Hits$ $taken$, $Penalties$, $Time$ $on$ $ice$, $Age$, and $Penalties$ $Taken$) cohorts. After matching, the standardized bias for each variable is less than 0.25 in both groups, and for most variables, the standardized bias is less the 0.10 among matched subjects. Given the reductions in bias after matching, it appears reasonable to consider the covariates' distributions balanced between the different treatment groups within the matched cohort.

Figures \ref{F1} - \ref{F4} also highlight the potential weaknesses of relying on regression-based strategies, such as multiple linear regression or a difference-in-differences design. More than half of the covariates' bias in the full sets of each of the North American and European players are greater than 0.25, which is problematic because differences in the distributions of $\boldsymbol{X}$ between treatment groups can lead to biased regression-based estimates when the regression model is misspecified \citep{cochran1973controlling, rubin2001using}. Further, there is a large subset of North American players with near-zero probabilities of playing abroad; attempting to identify the effect of playing abroad on players who may not have actually been able to play abroad, as would be done using a regression approach, requires extrapolation and making unjustifiable assumptions.  Finally, in general, regression-based strategies are more sensitive to covariate choice and model specification when compared to matched designs, and perform poorly in areas with limited covariate overlap \citep{dehejia1998causal, stuart2010matching}.

\subsection{Analysis phase}

We chose three measures to judge a player's performance after the lockout; his goals per 60 minutes, his assists per 60 minutes, and his $Shot \%$ (Corsi percentage). We also investigated a player's performance in both the short run (one month since the season resumed) and long run (the entire resumed season) because one may expect the performance gap to be short lived.

Differences in outcomes are contrasted within the matched cohort. Under unconfoundedness, these comparisons provide unbiased estimates of the effect of playing overseas on on NHL players' outcomes. Table \ref{Taboutcomes} shows the treatment effects for our three outcome variables in both the short run and the long run. We used Abadie-Imbens standard errors for each treatment effect, which helps account for the uncertainty in the matching procedure \citep{abadie2006large}.

Our results indicate limited support for enhanced productivity as a result of playing overseas during the lockout. In particular, we find that for European players, those who chose to play overseas during the lockout had significantly more goals per 60 minutes ($p-value < 0.01$) during the shortened 2012-13 season than those who did not. In our cohort, this difference represented between roughly 2 and 3 more goals per season among the Europeans that chose to play abroad.

We do not observe similar significant productivity gain (in terms of goals per 60 minutes) among European players in the short run. One reason could be because the season resumed on short notice, which can result in high transition cost for those who played in Europe since they would need to almost immediately report to the team and adjust to the new season in North America. For all the other outcome variables, there is no evidence of a significant positive or negative treatment effect of playing overseas on player outcomes regardless of the time frame or ethnicity.

\begin{table*}[htbp]
\centering
\caption{$ATT$ of playing abroad}
\begin{tabular}{l l l l l l} \toprule
\footnotesize
\label{Taboutcomes}
   &\multicolumn{2}{c}{North Americans ($ATT_{NA}$)} & & \multicolumn{2}{c}{Europeans ($ATT_{E}$)} \\
\hline
 Variable & One Month & Full Season & & One Month & Full Season\\
\hline
$Goals/60$	& -0.021 (0.067) & -0.031 (0.060)	& & 	0.080 (0.072) & 0.158 (0.057)***	\\
$Assists/60$	&-0.027 (0.123) & -0.019 (0.087) 	& &	-0.154 (0.119) & 0.007 (0.941)	\\
$Shot \%$		&-0.049 (1.387) & -0.447 (1.116)	& &   -1.621 (1.786) & -1.787 (1.604)	\\ \hline
\multicolumn{2}{c}{Standard errors in parentheses} \\
\multicolumn{2}{c}{*$p$ \textless 0.10, **$p$ \textless 0.05, ***$p$\textless 0.01}  \\
\bottomrule
\end{tabular}
\end{table*}

\section{Discussion}
The matching results indicate that playing overseas during the 2012-13 NHL lockout did not consistently offer a competitive advantage, which suggests that player behaviors during the lockout may not have systematically affected individual players' performance. While the results are consistent with findings from select studies such as \citet{lee_rupp2007}, they disagree with other prior studies on labor disputes and productivity such as \citet{mas2006} and \citet{kruger_mas2004}. We take into account the unique aspects of the 2012-13 NHL lockout and propose the following four possible explanations that are in line with our conceptual framework focusing on the impact of effort and experience on player productivity.

First, much of the previous literature on labor strife and productivity considers strikes instead of lockouts. Because strikes are initiated by workers while lockouts are imposed by employers, workers are more unprepared and have less control over the situation under lockouts, whereas if the workers are able to strike, then they may also intentionally shirk to exert more pressure toward the employers. And in the case of NHL, since the lockout was imposed before the season started, players also carried more of the financial burden - the lockout resulted in a significant financial loss to the players since an NHL player's annual salary is often on a million-dollar scale. The temporary but hefty financial loss together with the feeling of lack of control over the negotiation may actually motivate \emph{all} players to exert more effort into their extended off-season, regardless of whether they chose to engage in training domestically or playing competitively overseas, though effort may be more costly for those who stayed in North America\footnote{Besides the differences in compensations, those who played aboard may also find it easier to maintain and improve competitive conditioning due to the zero-sum nature of hockey games and competitiveness of European leagues.}. Studies, e.g. \citet{lee_rupp2007}, on industries where employees collect large labor market rents also find limited support for reduced effort and productivity among commercial airline pilots when facing pay cuts\footnote{The scale of labor market rent in occupations such as commercial airline pilots is of course not comparable to professional sports players.}.

Secondly, the lockout lasted only three months, which may not be enough time for sizable productivity difference to form\footnote{It is also possible that the veteran players are used to keeping up their conditions during the annual off-season. A short lockout may be viewed by some as an extended off-season, and it may not influence a player's competitiveness if he follows his routine off-season training schedule.}. Moreover, there was also wide expectation that owners and players would learn from the previous lockout and the lockout would resolve in a timely manner because of the tremendous financial loss caused by the previous lockout \citep{lockout2012}. Such expectation and hope could alleviate some of the uncertainty that players feel about the future and help them focus on improving their competitive strength. Since players expected that the season could resume at any given moment, those who chose not to play overseas would exert as much effort in their training as those who played overseas to be ready for in-game competition.

Furthermore, the advantage of playing more competitive games in European leagues may be offset by the high transaction costs and heterogeneity in competitiveness. Besides new teammates, coaches and fans, there are other nontrivial differences between European games and NHL games, which require significant and costly adjustment from the players. For instance, European games follow a different rink size\footnote{The NHL rink is 200 feet by 85 feet with goal lines being 11 feet from the end boards, while in most European leagues, the ice rink is 210 feet by 98 feet with goal lines being 13 feet from the end boards.}, rules (Olympic hockey rules) and playing styles. The fact that we find limited productivity gains for European players but no impact for North American provides further evidence to this argument since it is presumably easier for European players to adjust to European leagues due to their background and experience. Long-distance travel between North America and Europe can also negatively affect athletic conditions, especially when the new season resumed on short notice. Despite being the most competitive option outside the NHL, few European leagues can match the NHL in terms of competitiveness and talents, and there is strong heterogeneity in competitiveness among the European leagues.

Lastly, those who played overseas may not perceive European leagues as their long term destinations, and thus may not exert full effort during their stay in Europe. Such hesitation could be due to the relatively weak financial offers from European clubs\footnote{For instance, Russia's KHL, one of the highest paying leagues in Europe, limited the compensation of locked-out NHL player to no more than 65 percent their NHL salary (pro rata). And there are reports of players who took nominal compensation or were paid on a per-game basis for their games in Europe. On the other hand, the NHL Players' Association offered its members an escrow payment in October that was worth 8\% of previous year's salary, which already surpassed many of the short term or per-game based contracts offered by European clubs.} and fear of injuries that may affect their NHL career\footnote{Adding to these factors is also the fact that player insurance provided by the NHL teams does not cover games played outside the NHL.}. Because most of the players who played overseas during the lockouts are European players, it is possible that some of them may also have chosen to play in Europe for non-competitive reasons - it may simply be because these players would prefer to be close to home during the lockout just as most American and Canadian players chose to stay domestically during the lockout.

\section{Conclusion}

In this paper, we utilize a propensity score matching technique to present the first set of evidence on the impact of labor supply decisions during labor strife and work stoppage on worker productivity in the context of professional hockey players. In particular, we consider a unique natural experiment from the 2012-13 NHL lockout, during which approximately 200 players decided to play overseas while the rest stayed in North America and practiced on their own. We find limited evidence of enhanced productivity only among European players who chose to play abroad during the lockout, and no evidence of a benefit or drawback for North American players. This suggests that playing overseas during the lockout did not systematically improve the post-lockout player productivity, from which we can infer that player behaviors during the NHL lockout did not consistently affect a player's performance. Possible explanations for the lack of consistent productivity changes include: $(1)$ lockouts (as opposed to strikes) could actually create incentives for \emph{all} players to exert effort; $(2)$ the 2012-13 NHL lockout ended quite early with timely resolution already expected based on past lockout experience; $(3)$ European games may differ from the NHL ones in terms of intensity, setup, and rules; $(4)$ there may be hesitation to exert full effort while playing abroad due to weak financial incentives and injury concerns. While our results may not fully agree with previous empirical evidence from manufacturing and public sectors, our findings are consistent with literature on labor conflicts and worker performance in industries where employees collect large labor market rents, e.g. commercial airline pilots.

While we made use of the most recent NHL lockout, a related analysis would contrast hockey players before and after the entire 2004-05 season was lost. However, not only would it be difficult to ascertain information about which players chose to play abroad during this period, such an analysis could be compromised by the myriad changes to league policies made before the start of the 2005-06 season, including the implementation of the shootout, which were designed to increase game scoring and excitement. \cite{franck2012one} and \cite{lopez2013inefficiencies}, for example, identified changes in game outcomes and when comparing behavior before and after the 2004 lockout. In this respect, the analysis of the 2012 lockout would be preferred; when NHL play resumed after in January of 2013, it did so with, by and large, the identical set of rules and incentives to the previous season. A related area could consider the effects of mid-season stoppages in play, although, admittedly, these are rare in professional sports.

Within the field of professional sports, our study suggests that playing in alternative leagues in the event of league lockouts may not offer a systematic advantage over self-motivated practices especially if the lockouts are expected to be short-lived. On a broader note, with the decline of union representation and increasing empowerment to employers in many industries, it is important to understand the effect of lockouts as an alternative form of work stoppage during labor disputes, particularly in terms of its impact on labor productivity. Much of previous literature on work stoppage and labor productivity finds that worker productivity suffers as a result of strike. In contrast, our results imply that labor outcomes may not be affected under lockouts. One channel that helps explain the difference is that strikes and lockouts can create different incentive schemes for the parties involved. A broader policy implication of our findings is that consumer welfare may not suffer as much after a lockout compared to a strike due to the relatively unaffected post-lockout worker productivity and resulting product quality. Since the ability to impose a lockout can offer a significant bargaining advantage, lockouts are justifiably restricted in order to protect workers' welfare in the collective bargaining process. In light of the increasing percentage of lockouts in recent labor stoppage incidents, policy makers thus need to evaluate the welfare tradeoffs between employees and consumers when considering legislative changes to either form of work stoppage.


\pagebreak%

\nocite{*}
\bibliographystyle{itaxpf}
\bibliography{mybib}

\begin{thebibliography}{39}
\newcommand{\enquote}[1]{``#1''}
\providecommand{\natexlab}[1]{#1}

\bibitem[{Abadie and Imbens(2006)}]{abadie2006large}
Abadie, A. and G.~W. Imbens.
\newblock \enquote{Large Sample Properties of Matching Estimators for Average
  Treatment Effects.}
\newblock \emph{Econometrica}, 74(1), (2006), 235--267.

\bibitem[{Ahmed et~al.(2006)Ahmed, Husain, Love, Gambassi, Dell'Italia,
  Francis, Gheorghiade, Allman, Meleth, and Bourge}]{ahmed2006heart}
Ahmed, A., A.~Husain, T.~E. Love, G.~Gambassi, L.~J. Dell'Italia, G.~S.
  Francis, M.~Gheorghiade, R.~M. Allman, S.~Meleth, and R.~C. Bourge.
\newblock \enquote{Heart Failure, Chronic Diuretic Use, and Increase in
  Mortality and Hospitalization: An Observational Study Using Propensity Score
  Methods.}
\newblock \emph{European Heart Journal}, 27(12), (2006), 1431--1439.

\bibitem[{Akerlof(1982)}]{Akerlof1982}
Akerlof, G.~A.
\newblock \enquote{Labor Contracts as Partial Gift Exchange.}
\newblock \emph{The Quarterly Journal of Economics}, 97(4), (1982), 543--569.

\bibitem[{Austin(2009)}]{austin2009balance}
Austin, P.
\newblock \enquote{Balance Diagnostics for Comparing the Distribution of
  Baseline Covariates Between Treatment Groups in Propensity-Score Matched
  Samples.}
\newblock \emph{Statistics in Medicine}, 28, (2009), 3083--3107.

\bibitem[{Bairam et~al.(1990)Bairam, Howells, and Turner}]{bairam1990}
Bairam, E.~I., J.~M. Howells, and G.~M. Turner.
\newblock \enquote{Production Functions in Cricket: the {A}ustralian and {N}ew
  {Z}ealand Experience.}
\newblock \emph{Applied Economics}, 22(7), (1990), 871--879.

\bibitem[{Berri and Brook(2010)}]{berri2010evaluation}
Berri, D.~J. and S.~L. Brook.
\newblock \enquote{On the Evaluation of the ``Most Important" Position in
  Professional Sports.}
\newblock \emph{Journal of Sports Economics}, 11(2), (2010), 157--171.

\bibitem[{Bewley(1999)}]{bewley1999}
Bewley, T.
\newblock \emph{Why Wages Don't Fall During a Recession}.
\newblock Cambridge, MA: Harvard University Press, 1999.

\bibitem[{Chandrasekher(2013)}]{chandrasekher2013}
Chandrasekher, A.
\newblock \enquote{Police Labor Unrest and Lengthy Contract Negotiations: Does
  Police Misconduct Increase with Time Spent Out of Contract?}
\newblock Working paper, University of California, Davis - School of Law, 2013.

\bibitem[{Coates and Humphreys(2001)}]{coates_humphreys2001}
Coates, D. and B.~R. Humphreys.
\newblock \enquote{The Economic Consequences of Professional Sports Strikes and
  Lockouts.}
\newblock \emph{Southern Economic Journal}, 67(3), (2001), pp. 737--747.

\bibitem[{Cochran and Rubin(1973)}]{cochran1973controlling}
Cochran, W.~G. and D.~B. Rubin.
\newblock \enquote{Controlling Bias in Observational Studies: A Review.}
\newblock \emph{Sankhy{\=a}: The Indian Journal of Statistics, Series A},
  417--446.

\bibitem[{De~Paola and Scoppa(2011)}]{DePaola2011}
De~Paola, M. and V.~Scoppa.
\newblock \enquote{The Effects of Managerial Turnover: Evidence From Coach
  Dismissals in Italian Soccer Teams.}
\newblock \emph{Journal of Sports Economics}.

\bibitem[{Dehejia and Wahba(1998)}]{dehejia1998causal}
Dehejia, R. and S.~Wahba.
\newblock \enquote{Causal Effects in Non-Experimental Studies: Re-Evaluating
  the Evaluation of Training Programs.}
\newblock Tech. rep., National Bureau of Economic Research, 1998.

\bibitem[{Franck and Theiler(2012)}]{franck2012one}
Franck, E. and P.~Theiler.
\newblock \enquote{One for Sure or Maybe Three: Empirical Evidence for Overtime
  Play from a Comparison of {Swiss} Ice Hockey and the {NHL}.}
\newblock \emph{Jahrb{\"u}cher f{\"u}r National{\"o}konomie und Statistik},
  210--223.

\bibitem[{Gustafson et~al.(1999)Gustafson, Hadley, and Ruggiero}]{gustfson2000}
Gustafson, E., L.~Hadley, and J.~Ruggiero.
\newblock \enquote{Alternative Econometric Models of Production in {Major
  League Baseball}.}
\newblock In \emph{Sports Economics: Current Research}, edited by E.~G.
  J.~Fizel and L.Hadley. Westport, CT: Praeger, 1999.

\bibitem[{Hall(2004)}]{hall2004}
Hall, P.~V.
\newblock \enquote{`{W}e'd Have to Sink the Ships': Impact Studies and the 2002
  West Coast Port Lockout.}
\newblock \emph{Economic Development Quarterly}, 18(4), (2004), 354 -- 367.

\bibitem[{Holland(1986)}]{holland1986statistics}
Holland, P.
\newblock \enquote{Statistics and Causal Inference.}
\newblock \emph{Journal of the American Statistical Association}, 945--960.

\bibitem[{Kahane(2005)}]{kahane2005}
Kahane, L.
\newblock \enquote{Production Efficiency and Discriminatory Hiring Practices in
  the {National Hockey League}: A Stochastic Frontier Approach.}
\newblock \emph{Review of Industrial Organization}, 27(1), (2005), 47--71.

\bibitem[{Kahane et~al.(2013)Kahane, Longley, and Simmons}]{kahane2013}
Kahane, L., N.~Longley, and R.~Simmons.
\newblock \enquote{Returns to Thuggery in the National Hockey League: The
  Effects of Increased Enforcement.}
\newblock In \emph{The Econometrics of Sport}, edited by S.~K.
  Placido~Rodriguez and J.~Garci, 81--98. Northampton, MA: Edward Elgar
  Publishing, 2013.

\bibitem[{Kahn(2000)}]{Kahn2000}
Kahn, L.~M.
\newblock \enquote{The Sports Business as a Labor Market Laboratory.}
\newblock \emph{Journal of Economic Perspectives}, 14(3), (2000), 75--94.

\bibitem[{Krueger and Mas(2004)}]{kruger_mas2004}
Krueger, A.~B. and A.~Mas.
\newblock \enquote{{Strikes, Scabs, and Tread Separations: Labor Strife and the
  Production of Defective Bridgestone/Firestone Tires}.}
\newblock \emph{Journal of Political Economy}, 112(2), (2004), 253--289.

\bibitem[{Krumer et~al.(2014)Krumer, Rosenboim, and Shapir}]{Krumer2014}
Krumer, A., M.~Rosenboim, and O.~M. Shapir.
\newblock \enquote{Gender, Competitiveness, and Physical Characteristics:
  Evidence From Professional Tennis.}
\newblock \emph{Journal of Sports Economics}.

\bibitem[{Lee and Rupp(2007)}]{lee_rupp2007}
Lee, D. and N.~G. Rupp.
\newblock \enquote{Retracting a Gift: How Does Employee Effort Respond to Wage
  Reductions?}
\newblock \emph{Journal of Labor Economics}, 25(4), (2007), pp. 725--761.

\bibitem[{Lopez(2013)}]{lopez2013inefficiencies}
Lopez, M.~J.
\newblock \enquote{Inefficiencies in the {National Hockey League} Points System
  and the Teams That Take Advantage.}
\newblock \emph{Journal of Sports Economics}, 1527002513486654.

\bibitem[{Macdonald(2012)}]{macdonald2012adjusted}
Macdonald, B.
\newblock \enquote{Adjusted Plus-Minus for NHL Players using Ridge Regression
  with Goals, Shots, Fenwick, and Corsi.}
\newblock \emph{Journal of Quantitative Analysis in Sports}, 8(3).

\bibitem[{Mas(2006)}]{mas2006}
Mas, A.
\newblock \enquote{Pay, Reference Points, and Police Performance.}
\newblock \emph{The Quarterly Journal of Economics}, 121(3), (2006), 783--821.

\bibitem[{Rosenbaum and Rubin(1983)}]{rosenbaum1983central}
Rosenbaum, P.~R. and D.~B. Rubin.
\newblock \enquote{The Central Role of the Propensity Score in Observational
  Studies for Causal Effects.}
\newblock \emph{Biometrika}, 70(1), (1983), 41--55.

\bibitem[{Rubin(2001)}]{rubin2001using}
Rubin, D.
\newblock \enquote{Using Propensity Scores to Help Design Observational
  Studies: Application to the Tobacco Litigation.}
\newblock \emph{Health Services and Outcomes Research Methodology}, 2(3),
  (2001), 169--188.

\bibitem[{Schmidt and Berri(2004)}]{schmidt_berri2004}
Schmidt, M.~B. and D.~J. Berri.
\newblock \enquote{The Impact of Labor Strikes on Consumer Demand: An
  Application to Professional Sports.}
\newblock \emph{The American Economic Review}, 94(1), (2004), pp. 344--357.

\bibitem[{Schuckers et~al.(2011)Schuckers, Lock, Wells, Knickerbocker, and
  Lock}]{schuckers2011national}
Schuckers, M.~E., D.~F. Lock, C.~Wells, C.~Knickerbocker, and R.~H. Lock.
\newblock \enquote{National hockey league skater ratings based upon all on-ice
  events: An adjusted minus/plus probability (AMPP) approach.}
\newblock \emph{Unpublished manuscript. Available at http://myslu. stlawu.
  edu/\~{} msch/sports/LockSchuckersProbPlusMinus113010. pdf}.

\bibitem[{Scully(1974)}]{Scully1974}
Scully, G.~W.
\newblock \enquote{Pay and Performance in {Major League Baseball}.}
\newblock \emph{American Economic Review}, 64(6), (1974), 915--30.

\bibitem[{Sekhon(2008)}]{sekhon2008multivariate}
Sekhon, J.~S.
\newblock \enquote{Multivariate and Propensity Score Matching Software with
  Automated Balance Optimization: the Matching Package for R.}
\newblock \emph{Journal of Statistical Software, Forthcoming}.

\bibitem[{Splawa-Neyman et~al.(1990 [1923])Splawa-Neyman, Dabrowska, and
  Speed}]{splawa1990application}
Splawa-Neyman, J., D.~Dabrowska, and T.~Speed.
\newblock \enquote{On the Application of Probability Theory to Agricultural
  Experiments. Essay on Principles. Section 9.}
\newblock \emph{Statistical Science}, 5(4), (1990 [1923]), 465--472.

\bibitem[{Staudohar(2013)}]{lockout2012}
Staudohar, P.~D.
\newblock \enquote{The Hockey Lockout of 2012-2013.}
\newblock \emph{Monthly Labor Review}, 7, (2013), 1--10.

\bibitem[{Stuart(2010)}]{stuart2010matching}
Stuart, E.
\newblock \enquote{Matching Methods for Causal Inference: A Review and a Look
  Forward.}
\newblock \emph{Statistical science: a review journal of the Institute of
  Mathematical Statistics}, 25(1), (2010), 1.

\bibitem[{Todd and Wolpin(2007)}]{Todd_Wolpin2006}
Todd, P.~E. and K.~I. Wolpin.
\newblock \enquote{The Production of Cognitive Achievement in Children: Home,
  School, and Racial Test Score Gaps.}
\newblock \emph{Journal of Human Capital}, 1(1), (2007), 91--136.

\bibitem[{Turner(2013)}]{turner2013}
Turner, A.
\newblock \enquote{Labor Strife in Public Schools: Does it Affect Education
  Production?}
\newblock Working papers, Carnegie Mellon University, 2013.

\bibitem[{von Allmen et~al.(2015)von Allmen, Leeds, and Malakorn}]{Allmen2015}
von Allmen, P., M.~Leeds, and J.~Malakorn.
\newblock \enquote{Victims or Beneficiaries? Wage Premia and National Origin in
  the {N}ational {H}ockey {L}eague.}
\newblock \emph{Journal of Sport Management}.

\bibitem[{Winfree and Fort(2008)}]{winfree_fort2008}
Winfree, J. and R.~Fort.
\newblock \enquote{Fan Substitution and the 2004-2005 {NHL} Lockout.}
\newblock \emph{Journal of Sports Economics}, 425(9), (2008), 425--434.

\bibitem[{Zimbalist(2000)}]{zimbalist2000}
Zimbalist, A.
\newblock \enquote{Economic Issues in the 1998-1999 {NBA} Lockout and the
  Problem of Competitive Balance in Professional Sports.}
\newblock In \emph{The Economics of Sports}, edited by W.~S. Kern, 93--113.
  Kalamazoo, MI: W.E. Upjohn Institute for Employment Research, 2000.

\end{thebibliography}
\appendix

\end{document}